\documentclass[preprint2]{aastex}
\def\12{{1\ov 2}}

\def\ov{\over}

\def\lsim{\ \raise -2.truept\hbox{\rlap{\hbox{$\sim$}}\raise5.truept
\hbox{$<$}\ }}

\def\gsim{\ \raise -2.truept\hbox{\rlap{\hbox{$\sim$}}\raise5.truept
\hbox{$>$}\ }}

\def\mincir{\ \raise -2.truept\hbox{\rlap{\hbox{$\sim$}}\raise5.truept
\hbox{$<$}\ }}
\def\magcir{\ \raise -2.truept\hbox{\rlap{\hbox{$\sim$}}\raise5.truept
\hbox{$>$}\ }}

\shorttitle{Supermassive Black Holes in Galactic Nuclei}
\shortauthors{Cavaliere \& Vittorini}
\slugcomment{ApJ in press.}

\begin{document}


\title{Supermassive Black Holes in Galactic Nuclei}


\author{A. Cavaliere \and V. Vittorini}
\affil{Astrofisica, Dip. Fisica Univ. Tor Vergata
Roma,I-00133,Italy \\ApJ in press}




\begin{abstract}
We discuss the link between the observations of distant quasars
and those of massive dark objects in the cores of many local
galaxies. We show how the formation of early black holes gives
rise to the luminosity function of high $z$ quasars, while it
imprints into their dark local relics a related shape of the
mass-dispersion correlation. We propose that in its lower section
the correlation slope will tell the (otherwise uncertain) strength
of the feedback effects from the quasar radiation on the host
galaxies.
\end{abstract}




\section{Introduction}

Two news have recently kindled the field of the quasars and active
galactic nuclei.

The first concerns the farthest objects. Not only single quasars
(QSs) have been detected at redshifts out to $z=6.28$, but also
the statistics at $z \approx 5$ has improved to the point of
outlining the bright section of the luminosity function (Fan et
al. 2001a, 2001b).

These findings confirm that the population of the optically
selected objects goes through the most sharp and non-monotonic of
evolutions. The comoving  density of the bright sources rises on
the scale of a few Gyrs from the Bang, to peak at around $z
\approx 3$ (Shaver et al. 1996); later on it turns over and  falls
by factors $10^{-2}$  toward us  (Boyle et al. 2000). A similar
message comes from the radio band,  see Jackson \& Wall (1999).

The second piece of news concerns the cores of many local
galaxies, where massive dark objects (MDOs) ranging from a few
$10^6$ to a some $10^{9}\,  M_{\odot}$ had been detected within
regions from a few to tens of pcs, see Richstone et al. (1998).
Now (Ferrarese \& Merritt 2000; Gebhardt et al. 2000) such masses
are found to correlate tightly  with the velocity dispersion in
the body of the surrounding host galaxies.

The MDO masses fit into the framework provided by the long
standing arguments (see Lynden-Bell 1969) that indicate
gravitational contraction rather than thermonuclear burning to be
the dominant source of QS output. We just recall this conclusion
to hinge upon the high bolometric power $L > 10^{45}$ erg s$^{-1}$
of many such sources, and on the high compactness with sizes down
to $R \sim 10^{15}$ cm indicated for some of them; it also
requires an overall efficiency up to $\eta \sim 10^{-1}$ for
conversion of gravitational into radiative power.

But then the argument may be carried on to evaluating the masses
involved (Cavaliere et al. 1983); in terms of $ M_8 = M/10^8 \,
M_{\odot}$ these read
$$ M_8 \approx 3
\, (L_{45} \, \Delta t_{-1} \, R_{15} /  \eta_{-1})^{1/2} ~, \eqno
(1)$$
where we have used $L_{45} = L/ 10^{45}$ erg s$^{-1}$,  $R_{15} =
R/10^{15}$ cm, and the source life time $\Delta t= \Delta
t_{-1}\,10^{-1}$ Gyr. Such masses are consistent with the MDO
observations.

Here we intend to link the largest values of $M$ with the early
QSs, and to discuss how such a relation is to be extended into the
range of lower redshifts and smaller masses.
\section{Black holes and their environment}

The accreting black hole (BH) paradigm not only best accounts for
the small sizes and the top efficiencies required, but also
provides to the spent masses the stability of a terminal
configuration (Rees 1984). Nailing down the indication from eq.
(1), a BH keeps full record of the mass
$$ M =
\int dt\, L(t) / \eta\,c^2 ~\eqno(2)$$
accreted over the life of an active galactic nucleus.

Meanwhile, the induced bolometric luminosities
$$L \approx \eta\, c^2 \Delta m/\Delta t ~ \eqno(3)$$
are tuned (even at constant $\eta \sim 10^{-1}$) by the mass
$\Delta m$ accreted over a time $\Delta t$, and cover a wide
dynamic range: from Eddington {\it self-limiting} conditions
governed by radiation pressure that yield $L \sim L_E \approx
10^{46}\, M_8$ erg s$^{-1}$, to {\it supply-limited} accretion
(Cavaliere \& Padovani 1989) that easily allows only sub-Eddington
emission.

Thus strong gravity is not enough. Equally important are the {\it
environmental} conditions in, or surrounding the host galaxy;
these can drive widely ranging accretion rates from $10^{-3}$ or
less, up to some $ 10^2 \, M_{\odot}$ yr$^{-1}$. In addition, it
is the cosmological change of the environment that qualifies to
govern the QS evolution; in fact, both occur on time scales
$t_{ev}\sim$ a few Gyrs such that $\eta\, t_E\, \ll \, t_{ev}\,\ll
\, H_0^{-1}$ holds.

In turn, the environmental conditions are described by the other
paradigm, the hierarchical growth and clustering of dark matter
(DM) halos, wherein the galaxies constitute lighter baryonic cores
(White \& Rees 1978). This implies substantial dynamical events to
occur, at early $z$ in the strong form of merging between
comparable subgalactic units, later on as milder interactions of
galaxies in groups.

All these dynamical events tend to break on scales of kpcs the
axial symmetry of the galactic gravitational potential, or enhance
its steady  asymmetry; relatedly, the specific angular momentum
$j$  providing support to the gas in the central kpc of the host
is not conserved, rather it is transferred to the massive DM
component. Thus the necessary condition is provided for
destabilizing and funneling inward a sizeable gas fraction. At
smaller scales dissipative processes take over to redistribute $j$
(Haehnelt \& Rees 1993), and cause the gas to reach the nuclear
accretion disk and grow new BHs or refuel the old ones.

To link BH growth and QS luminosities we follow  Cavaliere \&
Vittorini (2000) and disentangle  the triggering dynamical events
into two main {\it regimes}, roughly divided by the epoch of group
formation $z_G \simeq 2.5 \pm 0.5$, depending on
cosmological/cosmogonical parameters.

\section {QSs in forming spheroids for $z>$ 3}

The {\it self-limited} regime occurs  mainly  at epochs before
$z_G$, when galactic spheroids are built up through major merging
events between  halos with $M_h < 10^{13}\, M_{\odot}$. These
events destabilize large amounts of gas, but also replenish the
host structures with fresh supplies and sustain the gas amount at
the cosmic level $m \approx 10^{-1}\, M_h$.

As a consequence, central BHs can form and/or accrete rapidly,
growing by  $\Delta m \sim M$ over dynamical times close to the
Salpeter scale, $t_{dyn} \sim \eta t_E$; so after eq. (3) they
attain Eddington luminosities  $L\sim L_E \propto M$.

In turn, $M$ is related to $M_h$; two specific models bracket the
processes involved, see Cavaliere \& Vittorini 1998 (CV98),
Hosokawa et al. (2001). Haehnelt \& Rees (1993) considered BH
coalescence in parallel with merging of their halos; this process
(hereafter model A) is described by the simple scaling
$$ M \approx 10^{-4} \, M_h ~. \eqno (4 A) $$

Alternatively, Haehnelt, Natarajan \& Rees (1998) proposed the
feedback-constrained model B where the scaling reads
$$ M_8 \approx
(1+z)^{5/2}\, M_{h,13}^{5/3} ~.  \eqno(4 B)$$
This is because during halo merging a central BH may also accrete
gas up to the limit
$$\epsilon \, L_E\, t_{dyn}  \mincir  G\,M_h\, m /r ~;     \eqno(5)$$
this is set by gas unbinding from the halo potential well, due to
the deposition of an (uncertain) fraction $\epsilon  \sim 10^{-2}$
of the QS radiation (Silk \& Rees 1998).

With both models (4A) and (4B) the early QSs are expected to grow
in average luminosity and in number, tracking the development of
protogalactic halos over the range from $M_h \sim 10^{10}$ toward
$10^{13}\, M_{\odot}$. The evolving halo mass distribution is
widely taken  in the form $N_{PS} (M_h, z)$ first proposed by
Press \& Schechter (1974); the positive term of its time
derivative provides the rate of halo formation, and yields (see
CV98) the luminosity function (LF) in the form
$$N(L,z)\;dL\; = \;\Delta t\;\partial _tN_{PS}(M_h,z)\;dM_h ~,    \eqno(6)$$
with the prefactor $\Delta t \approx \eta\, t_E$ accounting for
the limited source lifetime.

Fig. 1 shows the  optical LFs provided by such models. We stress
that model (4B) by the very means of its non-linear transformation
stretches the halo distribution into flatter, more fitting LFs and
predicts a stronger (negative) evolution. By the same token, it
also associates bright QSs with the largest galactic halos,
consistent with the data discussed by Hamilton, Casertano \&
Turnshek (2000).

Both models are normalized to the data at $z \approx 4$. But model
(4B) privileges the upper halo range where $N(M_h)$ decreases
steeply, so it provides BH numbers smaller and naturally close to
{\it one} large BH per actively star forming protogalaxy of
intermediate (Steidel et al. 1999) or large mass (Granato et al.
2001).

In either model, the early mass distribution $N (M, z)$ is
directly related to $N_{PS} (M_h, z)$; the model (4B) yields the
result represented by the thick solid line in fig. 2.

\section {QS$s$ in interacting galaxies for $z<3$}

After $z \approx z_G$ the galaxies are assembled into small groups
of mass $M_G \magcir 10^{13} \, M_{\odot}$, where the dominant
member recurrently interacts with its companions, to the effect of
refueling and rekindling an old BH; growing evidence (referenced
in CV00) relates many QSs with interacting hosts. Small groups,
with their high galaxies density and low velocity dispersion $V$
still close to the galaxian dispersion $\sigma$, constitute
preferred sites for such interactions to occur.

{\it Supply-limited} accretion prevails here, since now the gas
mass $m$ in the host is depleted but no longer replenished by the
interactions. Still, considerable fractions
of the gas initially orbiting in the host at $r \sim $  kpc are
destabilized and partly made available for accretion; such
fractions are
easily evaluated (see CV00) in the form
$$f_d \mincir \Delta j/j  \approx G\, M'/ \sigma \, V \, b ~.
\eqno (7) $$
This includes the host structural parameter $j/r$, taken to be
close to  $\sigma$; it also includes encounter orbital parameters:
the impact parameter $b$, the relative velocity $V$, and the
partner mass $M'$. Truly tidal interactions imply a postfactor
$r/b$.

With $V \magcir \sigma$ and $b$ bounded by the group radius,
interactions and gas inflow both take times $\Delta t $ $\approx
b/V \sim 10^{-1}$ Gyr. The above equation easily yields $f_d $ in
excess of a few $\%$, of which about $1/3$ may reach the nucleus
while the rest is likely to end up in circumnuclear starbursts,
see Sanders \& Mirabel (1996), CV00. This is enough to produce
(see eq. 3) outputs  $L \magcir 10^{46}$ erg $s^{-1}$ in a host
still gas rich with $m \sim 10^{10}\,  M_{\odot}$. Fractions up to
$f_d \approx 1/2$ are indicated by eq. (7), and in fact obtain in
numerical simulations of grazing collisions, which are
statistically fewer. The full probability of accreting a fraction
$f$ is calculated by CV00 to read $P(f) \propto f^{-2}$ on the
basis of eq. (7); this defines range and shape of the LF.

In time, many small groups  merge into richer ones where
interactions are less frequent and effective. Their volume
density, grown rapidly later than $z_G$, at low $z$ goes down into
a demise $N_G(z)\propto (1+z)$; in addition, the decreasing
interaction rate $\tau_r^{-1}(z) \approx 0.5\,
(1+z)^{3/2}$Gyr$^{-1}$ lowers the number of activated sources. The
result for the QS population is a moderate ``density evolution"
proportional to $N_G(z)\, \tau_r^{-1}(z)$. But a stronger
``luminosity evolution'' occurs since on average $L \propto f \,
m(z)$ holds, and the residual host gas $m(z)$ is depleted on a
scale $t_{ev}$ as it is destabilized and used up in accretion
episodes and by accompanying nuclear starbursts, with no
replenishment.

As $L$ decreases and $M$ increases, the sources on average go
toward sub-Eddington luminosities, consistent with the data in
Salucci et al. (1999), Wandel (1999).

\section{Relics at $z \approx 0$}

For early BHs growing at $z > 3$ the models (4A) or (4B) imply
different scaling relations, that read
$$M \propto  \sigma^3 \, \rho^{-1/2}(z) \propto \sigma^4 ~, \eqno (8 A)$$
or
$$M \propto  \sigma^5 ~, \eqno (8 B)$$
respectively. These stem from the simple hierarchical scaling of
$M_h \propto V^2\, R/ G$ and $R \propto$ $(M_h/\rho)^{1/3}$ for
the surrounding DM halos, as is appropriate for $M_h < $
$10^{13}\, M_{\odot}$ when one galaxy per halo occurs. In such
conditions, it is also fair to assume
 $\sigma  \propto $ $V$, and to fix
the $z$-dependent fuzz appearing only in eq. (8A) on relating the
density to $M_h$ by means of $\rho \propto M_h^{-1/2}$, that holds
for hierarchically formed halos (see Haehnelt \& Kauffmann 2000).

The relations (8A), (8B) turn out to be in tune with the current
debate concerning the MDO data. These have been recently
recognized to follow a tight and steep correlation; the precise
slope is given as slightly flatter than $M \propto \sigma^4$ by
Gebhardt et al. (2000), or close to $M \propto  \sigma^5$ by
Ferrarese \& Merritt (2000). In either case the scatter is found
to be rather small, that is, factors $10^{\pm 0.35}$ or less in
$M$ at given $\sigma$. Note from fig. 1 that LFs flat as observed
at high $z$ obtain from model (4A) only upon convolution with
scatter larger than $10^{\pm 0.5}$, as discussed by Haiman \& Loeb
(1998).

Later than $z \approx 3$ fewer new BHs are still produced, but the
early ones still grow after $M(z) =   M(3) + \sum_i \Delta m_i$,
due to additional but dwindling accretion events $\Delta m_i$
caused by host interactions.

The latter cause the mass distribution $N(M,z)$ to evolve as given
by
$$ \partial_t N =  {N_G\over N_{tot}\,
\tau_r} \, \int df P(f)\,  [N (M - fm) - N(M)]~. \eqno (9)$$
Following up \S 4, $N_G(z)/N_{tot}$ is the fraction of host
galaxies residing in a group, relative  to total  BH number
including the dormant ones; for it we take values of order
$10^{-1}$ as discussed in CV00, based on the fraction $1/3$ of
bright galaxies found by Ramella et al. (1999) to reside in groups
with membership 3 or larger. The right hand side describes the net
change at $z$ of the distribution $N(M,z)$, due to an accretion
episode of $\Delta m = f\, m (z)$ which upgrades the initial mass
$M-\Delta m$ to the current value $M$; the probability for this to
occur is $P(f)$, and the factor provided by the interaction
frequency $\tau^{-1}_r (z)$ converts it to a rate.

The numerical solution  plotted in fig. 2 shows how $N(M, z)$ --
starting from the condition at $z=3$ computed in \S 3 -- drifts
and diffuses in time toward higher $M$, but only up to a cutoff;
this is due to the large values of $f$ being rare, and to $m(z)$
being depleted from its initial value $10^{-1} M_h$. So the
additions $\Delta m = f\, m(z)$ decrease on average, and large $M$
are unlikely to grow much larger.

We derive the corresponding  $M-\sigma$ relation noting from eq.
(7) that the gas masses available for accretion follow $\Delta m =
f_d \, m/3 \propto  m / \sigma$. On adopting the scaling  $m
(\sigma) \propto \sigma^4$ indicated by the Faber-Jackson relation
or produced by stellar feedback, the result is $\Delta m \propto
\sigma^3$. If used in full, such masses dominate at low $\sigma <
200$ km s$^{-1}$, to yield maximal $M$ values scaling as
$$M \propto \sigma^3
~, \eqno (10)$$
see fig. 3. The scatter is within an overall factor $5$, the
effective range given by the steep form of $P(f)$. Truly tidal
interactions (see the comment under eq. 7) yield a somewhat
steeper scaling $\Delta m \propto  m (\sigma) \, r/\sigma \propto
\sigma^4$ and more scatter.


But -- given that enough gas is made available by the interactions
as shown in \S 4 -- the mass actually accreted  may still be {\it
constrained} by the QS feedback, depending on the  degree of
coupling of the source output to the surrounding baryons. Here we
focus on the 90\% radio quiet sources where the output is mostly
radiative and roughly isotropic. At low $z$ the unbinding
constraint similar to eq. (5) reads
$$ L \,
\Delta t\, \approx  \eta c^2\Delta m \mincir   \epsilon^{-1}\,
m(z)\, \sigma^2  ~. \eqno (11)$$
With $\epsilon \sim 10^{-2}$, the masses accreted in a host with
$\sigma $ $< 200$ km s$^{-1}$ are constrained to stay under those
made available by interactions and given by eq. (7); using again
$m\propto \sigma^4$, here the result is $\Delta m_i \propto
\sigma^{6}$. These moderate additions drive considerably less
evolution of $N(M, z)$ as represented by the thin solid line in
fig. 2; then $M$ is always dominated by the initial values which
follow eq. (8B), and will in fact converge to values scaling as $M
\propto \sigma^{5}$.

The full scaling laws given by eqs. (8B) with (11), or by (8A)
with (10), are represented in fig. 3. They differ mostly in the
lower left section, where data are difficult to obtain from the
kinematics of stars or of a nuclear disk.

\section{Conclusions and discussion}

We conclude that the relic BH masses are related to the host
galaxy dispersions by the {\it steep} correlation $M \propto
\sigma^5$ if QS feedback onto the host's gas always provides the
main constraint to the accretion. In other words, this form is to
hold at all $\sigma$ if the host potential well {\it bounds} the
actual accretion rates all the way from early BH formation to
later additional growth.

On the other hand, if such a control was never the limiting factor
we expect the softer (and fuzzier) correlation $M \propto
\sigma^4$ to hold at high $\sigma$, and this to {\it soften} yet
at lower $\sigma$ to the form $M \propto \sigma^3$ set by the
production rate of the available gas.

These two extremes {\it bracket} the QS and BH story. The latter
is centered on one kind of engine, the BH based on strong gravity;
but it comprises different regimes of fueling, that is, of
accretion and growth triggered by the environment under weak
gravity.

At early $z \magcir 3$ much fresh gas is made available for full,
self-limiting accretion by major merging events which build up the
halos and the embedded spheroids. BHs grow rapidly, generating the
upper section of the correlation $M \propto \sigma^4 \div
\sigma^5$. Meanwhile, the associated QSs flare up at Eddington
luminosities; their LF grows in range and height, tracking the
progressive halo build up.

Later than $z \approx z_G \approx 2.5$ the dearth of gas curbs the
accretion and bends down the QS evolution, as host interactions
with group companions are still able to trigger accretion but no
longer to import fresh gas. Though supply-limited, the gas masses
made available for accretion are still sizeable. If these are used
in full, the average luminosities are still high and the BHs still
grow considerably; relatedly, the correlation is softened to $M
\propto \sigma^3$ in its lower section.

At $z$ lower yet, the host gas reservoirs approach exhaustion, and
weaker AGN activity may be sustained by lesser gas productions or
smaller supplies provided, e.g., by internal instabilities (see
Heller \& Shlosman 1994, Merritt 1999) or by satellite galaxies
cannibalized by the hosts (CV00). These latest and smaller
accretion events can fuel many weaker AGNs more likely to be
pinpointed in X-rays, implying widespread accretion power but
moderate growth of individual BHs, as also argued by Salucci et
al. (1999). Deep surveys in hard X-rays are now uncovering
optically hidden, and even X-ray obscured accretion power, as they
resolve and count the AGNs that dominate the hard XRB (see
Giacconi et al. 2001, Hasinger et al. 2001); the indications are
currently favoring AGNs of intermediate powers and redshifts as
main  XRB contributors (see Comastri et al. 2001). These findings
jointly with the scanty evidence of, and the indirect bounds to a
QS 2 population (see Maiolino et al. 2001) suggest to us that the
optical-IR sampling is fair to within a factor 2 as for the
outputs of early and luminous QSs, and is reliable for linking
these with the largest BHs.

Thus our conclusions may be rephrased in terms of two simple {\it
patterns} standing out of this complex picture.

First, the largest MDOs in local inactive galaxies are directly
{\it linked} with the high-$z$ luminous QSs. In particular, we
have shown that {\it flat} optical LFs given by the
feedback-constrained model (4B) are linked to the {\it steep} and
{\it tight} correlation given at large masses by $M\propto
\sigma^5$ (see eq. 8B). Current evidence concerning high
brightness of the hosts (see \S 3), flat shape of the high-$z$ LFs
(\S 3), and narrow scatter in the $M-\sigma$ correlation (\S 5)
concur to favor this specific link. But telling data will be
provided by the precise slope of the upper MDO correlation
observed locally in the optical band, jointly with the rate of QS
decline toward high $z$ derived from large area surveys out to the
near-IR like the Sloan Digital Sky Survey.

Second and concerning the lower region of the  $M$--$\sigma$
plane, it is conceivable (within the present uncertainties about
$\epsilon$) that the fuel throttle to the BH engine is always
controlled by the host wells as is expressed by eq. (11); if so,
the main accretion modes may be effectively united by the feedback
action. In \S 5 we have shown that this provides the {\it lower}
bound to $M$ expressed again by $M \propto \sigma^5$. On the other
hand,  an {\it upper} bound $M \propto \sigma^3$ is set by the
condition of maximal fuel availability, eq. (10); this includes
the production of optically obscured accretion power. Note that
the highly obscured, wind blowing BHs suggested by Fabian (1999)
would be located in between. The actual feedback action  will be
observationally probed by the slope of the $M-\sigma$ correlation
in its lower section; here the growing reverberation map database
(see Ferrarese et al. 2001) in current AGNs will be telling.

\acknowledgements

\noindent We thank F. Bertola and L. Danese for  informative
discussions and helpful comments. Work supported by ASI and MIUR
grants.



\clearpage


\begin{figure}
\figurenum{1} \plotone{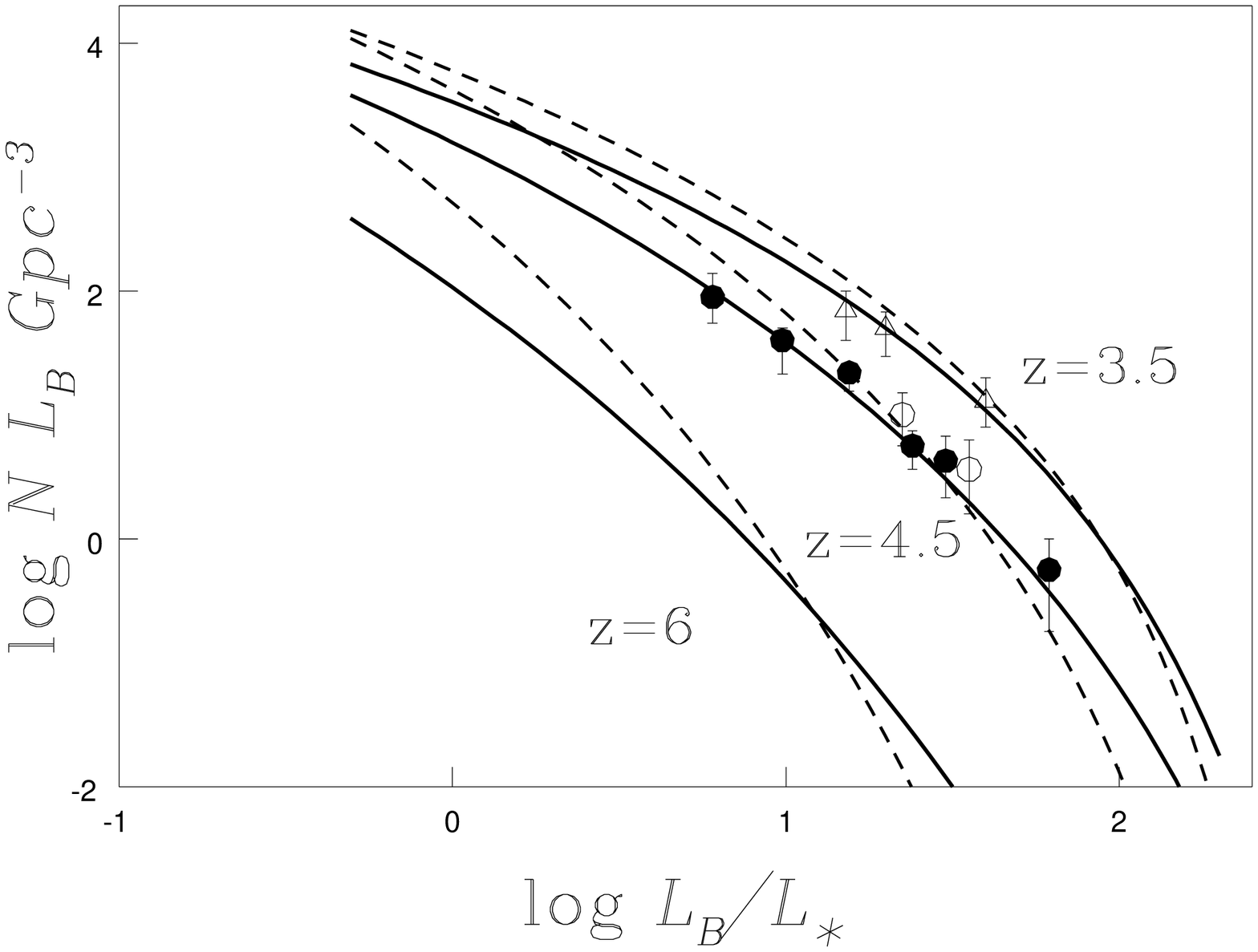} \caption{High-$z$ LFs from the
models discussed in \S 3: eq. (4A), dotted; eq. (4B), solid. Here
$L_B$ is the luminosity in the blue band (bolometric correction
$\kappa =10$), and $L_* = 10^{45}$ erg s$^{-1}$; canonical
$\Lambda$CDM cosmology/cosmogony. Data: filled circles from
Kennefick et al. 1995, Schmidt et al. 1995; open symbols from Fan
et al. 2001a. \label{fig1}}
\end{figure}


\begin{figure}
\figurenum{2} \plotone{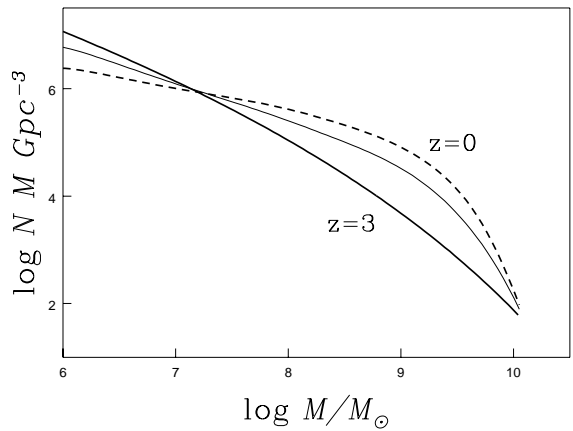} \caption{The evolution of the mass
distribution $N(M, z)$ of the relic BHs. Down to $z \approx 3$
this is given by  $N_{PS}(M_h ,z)$ with the transformation (4B),
see the thick solid line.  Afterwards, the distribution is driven
by host interactions as described by eq. (9). The dashed line
corresponds to unconstrained accretion of $1/3$ of the gas sent
toward the nucleus; the thin solid line corresponds to accretion
constrained by the QS feedback following eq. (11). The related
local mass density amounts to $\rho_{BH}\approx  4$ and $2\,
10^{14}\, M_{\odot}$ Gpc$^{-3}$, respectively. Cosmology/cosmogony
as in fig. 1. \label{fig2}}
\end{figure}

\begin{figure}
\figurenum{3} \plotone{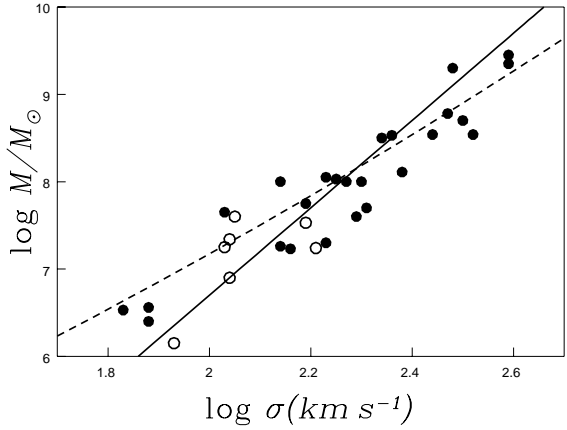} \caption{The models for the local
$M - \sigma$ relation discussed in \S 5. Dashed: unconstrained
accretion at levels $f= 5\%$, the average value given by the
distribution $P(f)$. Solid: feedback constrained accretion with
$\epsilon = 5 \, 10^{-3}$. Data from Gebhardt et al. 2000, and
Ferrarese \& Merritt 2000, with the open symbols marking results
from reverberation mapping. The model normalizations at high $M$
follow the fits by the same authors. \label{fig3}}
\end{figure}

\end{document}